\documentclass[11pt]{article}

\usepackage{amssymb}
\usepackage{amsmath}
\usepackage{hyperref}

\begin{document}

\title{On the Relationship between the Moyal Algebra and the Quantum Operator Algebra of von Neumann.}
\author{B. J. Hiley.}
\date{TPRU, Birkbeck, University of London, Malet Street,\\ London WC1E 7HX.\\ \vspace{0.2cm} [b.hiley@bbk.ac.uk]}
\maketitle

\begin{abstract}
The primary motivation for Moyal's approach to quantum mechanics was to develop a  phase space formalism for quantum phenomena by generalising the techniques of classical probability theory.  To this end, Moyal introduced a quantum version of  the characteristic function which immediately provides a probability distribution.  The approach is sometimes perceived negatively merely as an attempt to return to classical notions, but the mathematics Moyal develops is simply a re-expression of what is at the heart of quantum mechanics, namely the non-commutative algebraic structure first introduced by von Neumann in 1931.   In this paper we will establish this relation and show that  the ``distribution function'', $F(P,X,t)$ is simply the quantum mechanical density matrix for a single particle. The coordinates, $X$ and $P$, are not the coordinates of the particle but the mean co-ordinates of a cell structure (a `blob') in phase space, giving an intrinsically non-local description of each individual particle, which becomes a point in the limit to order $\hbar^2$.  We discuss the significance of this non-commutative structure on the symplectic geometry of the  phase space for quantum processes.  
 \end{abstract}


\section{Introduction.}
As is well known, the Wigner-Moyal approach has its origins in an early attempt of Wigner \cite{wig} to  find quantum corrections to the statistical properties of thermodynamic quantum systems.   A major contribution to the development of the formalism that will be the subject of this paper comes from the seminal paper of Moyal \cite{moy}.  The  aim of his paper was to
``reformulate the principles of quantum mechanics in purely statistical terms".   In doing this, he has constructed what looks like a  classical $(x,p)$ phase space approach to quantum phenomena even though quantum operators play a primary role in setting up the formalism.  His aim was to find a generalisation of the traditional methods of statistics to meet this challenge.  In fact what he has created is a non-commutative algebraic structure based on a symplectic symmetry, namely, the symplectic Clifford algebra \cite{ac90}.  In this paper I will discuss how this structure arises and forms a companion to the orthogonal Clifford algebra that plays a central role in the description of the Dirac electron.  Here we will discuss the role of the symplectic structure leaving the account of how these two algebras can be combined to a future paper \cite{soc12}

Moyal's work initially received considerable opposition from Dirac \cite{amoy}.  These objections and other misunderstandings have prevented the implications of Moyal's work from being fully appreciated in the physics community at large although it is used successfully in quantum many body scattering problems \cite{pcfz83}.  In spite of this success, one is left with the belief that it is some form of `semi-classical'  approach, which works well in many different situations discussed in \cite{pcfz83}.  

The terminology implies that it is essentially an attempt to return to a classical-type description.  However this is not correct as it is a full quantum description in it own right. To appreciate this fact, it is necessary to be aware that the mathematical structure used by Moyal is identical to the formalism developed by von Neumann in a very early paper ``Die Eindeutigkeit der Schr\"{o}dingerschen Operatoren" \cite{von}.  The content of this paper is significantly different from that of his classic book ``Mathematische Grundlagen der Quantenmechanik" \cite{vn32} to warrant reexamination.

  In this paper I want to bring out the exact relation between Moyal's approach \cite{moy} and this early work of von Neumann, not merely for historic reasons, but also to propose a new way forward to explore the  non-commutative geometric aspects of the Moyal algebra that does not put the wave function centre stage \cite{jgbjv95} \cite{hg83}.

\section{Moyal's Proposals.}

\subsection{The Characteristic Function}

To explore the possibility that the standard approach to quantum mechanics disguises a deeper statistical theory based on traditional statistical ideas, Moyal proposed that we start by defining one of the standard tools of a statistical theory, namely, the characteristic function.  To extend these statistical methods to the quantum domain, he introduced a {\em characteristic operator} namely, the Heisenberg-Weyl operator  
\begin{eqnarray}
{\widehat M}(\tau,\theta)=e^{i(\tau{\hat p}+\theta{\hat x})} 	\label{eq:1}
\end{eqnarray}									
where ${\hat x}$ and ${\hat p}$ are the usual quantum operators with $\tau$ and $\theta$ being arbitrary parameters. 
To define the characteristic function, we first take the expectation value of $\widehat M(\tau, \theta)$ by writing
\begin{eqnarray}
M_\psi(\tau,\theta,t)=\langle\psi|\widehat M(\tau,\theta)|\psi\rangle=\int \psi^{*}(x,t) e^{i(\tau {\hat p}+\theta{\hat x})}\psi(x,t)dx.	\label{eq:2}
\end{eqnarray}							
Using $e^{i(\tau\hat p+\theta\hat x)}=e^{-i\tau\hat p/2}e^{i\theta \hat x}e^{i\tau\hat p/2}$, expression (\ref{eq:2}) can immediately be reduced to	
\begin{eqnarray}
M_\psi(\tau,\theta,t)=\int\psi^{*}(x-\tau/2,t)e^{i\theta x}\psi(x+\tau/2,t)dx.
\label{eq:3}
\end{eqnarray}									
$M_\psi(\tau, \theta,t)$ is then defined to be the characteristic function.  As in classical statistics the probability distribution $F_\psi(p,x,t)$ can be found simply by taking the Fourier transform of $M_\psi(\tau,\theta,t)$ so that
 \begin{eqnarray}
F_\psi(p,x,t)&=&\frac{1}{(2\pi)^{2}}\int\int M_\psi(\tau,\theta,t)e^{-i(\tau p+\theta x)}d\tau d\theta\nonumber\\
&=& \frac{1}{2\pi}\int\psi^*(x-\tau/2,t)e^{-ip\tau}\psi(x+\tau/2,t)d\tau.\label{eq:4}										
\end{eqnarray}
This will immediately be recognised as the  distribution first introduced by Wigner \cite{wig}.  It should be noted that here $p$ and $x$ are continuous variables in some, as yet unspecified phase space and should not immediately be thought of as specifying the position and momentum of a single particle.  Rather it should be thought of as some  arbitrary coordinate system on a symplectic manifold with an invariant 2-form, $\sigma(p,x)$ yet to be defined\footnote{Here we are concerned with conceptual questions so the mathematical formalism will be kept as simple as possible using a two-dimensional phase space. The generalisation to a higher dimensional phase space is straight forward.}.

Finally we can find the expectation value of any bounded operator through
\begin{eqnarray}
\langle \hat A\rangle =\int a(p,x)F_\psi(p,x,t)dxdp		\label{eq:exptM}
\end{eqnarray}
where $a(p,x)$ is an ordinary function on the symplectic space.

  \subsection{The von Neumann Approach.}
  
 It is interesting to note that the expression (\ref{eq:2}), which Moyal called a `characteristic function', had already appeared in a classic paper by von Neumann, ``Die Eindeutigkeit der Schr\"{o}dingerschen Operatoren''  \cite{von}, but von Neumann gave it no name.  In his paper, von Neumann laid the foundations of what we  now know as  the Stone-von Neumann theorem, in which he showed that the Schr\"{o}dinger representation is irreducible and unique up to a unitary equivalence.  
 
Indeed what we will now show is that the mathematical structure developed by von Neumann is identical to the one that appeared in Moyal's paper.  Thus the Moyal algebra lies at the very heart of the standard operator formulation of quantum mechanics.   

Rather than starting with the well known relation 
\begin{eqnarray*}
[\hat x, \hat p]=i\hbar
\end{eqnarray*}
von Neumann introduces a pair of bounded operators, $U(\alpha)=e^{i\alpha {\widehat p}}$ and $V(\beta)=e^{i\beta{\widehat x}}$ so that the non-commutative multiplication can be written in the form
\begin{eqnarray}
U(\alpha)V(\beta) = e^{i\alpha\beta}V(\beta)U(\alpha).	\label{Weyl}
\end{eqnarray}
together with
\begin{eqnarray}
U(\alpha)U(\beta)=U(\alpha + \beta);  \hspace{0.5cm}  V(\alpha)V(\beta)=V(\alpha + \beta). \nonumber
\end{eqnarray}	
These relations were originally introduced by Weyl \cite{hw28}.								

von Neumann now defines an operator
\begin{eqnarray}
\widehat S(\alpha,\beta)=e^{-i\alpha\beta/2}U(\alpha)V(\beta)=e^{i\alpha\beta/2}V(\beta)U(\alpha)  \nonumber
\end{eqnarray}
which can also be written in the form
\begin{eqnarray}
\widehat S(\alpha, \beta)=e^{i(\alpha \hat p+\beta \hat x)}.	\label{eq:hesgen}		
\end{eqnarray}
This is exactly the operator $\widehat M(\tau, \theta)$ introduced in equation (\ref{eq:1}), only the notation is different.  Thus Moyal's starting point is  exactly the same  as that of von Neumann. 

Let us go further.  von Neumann then proves that the operator $\widehat S(\alpha, \beta)$ can  be used to define any bounded operator $\hat A$ on a Hilbert space  through the relation
\begin{eqnarray}
\hat A=\int\int a(\alpha, \beta)\widehat S(\alpha,\beta)d\alpha d\beta.
\label{eq:symA}				
\end{eqnarray}
where $a(\tau,\theta,t)$ is the kernel of the operator. 

To proceed further,  von Neumann defines the expectation value of the operator $\hat A$ as
\begin{eqnarray}
\langle \psi|\hat A|\psi\rangle = \int\int a(\alpha, \beta)\langle \psi|\widehat S(\alpha,\beta)|\psi\rangle d\alpha d\beta.  \label{eq:expA}		
\end{eqnarray}
Here 
\begin{eqnarray*}
\langle\psi|\widehat S(\alpha,\beta)|\psi\rangle=\langle\psi|e^{i(\alpha \widehat p+\beta\widehat x)}|\psi\rangle
\end{eqnarray*}
so that
\begin{eqnarray*}
\langle \psi|\widehat S(\alpha,\beta)|\psi\rangle=\langle \psi|\widehat M(\tau,\theta)|\psi\rangle,
\end{eqnarray*}
which, apart from a change of variables, is identical to the expression used by Moyal in equation (\ref{eq:2}).  If we now use the Fourier transformation of $M_\psi(\tau, \theta,t)$ in equation (\ref{eq:expA}), we find it immediately gives the Moyal equation (\ref{eq:exptM}) for the expectation value, which for an operator that is time dependent gives
\begin{eqnarray}
\langle\psi| \hat A(t)|\psi\rangle = \int\int a(p,x,t) F_\psi(p,x,t)dpdx.  \label{eq:expt}
\end{eqnarray}								
Here $a(p,x,t)$ is called the {\em Weyl symbol}, its Fourier transform being $a(\tau,\theta,t)$.

Moyal noticed that if $a(p,x,t)$ could be regarded as one of the possible values of $\widehat A$ and if we could regard $F_\psi(p,x,t)$ as a probability distribution, then the RHS of  (\ref{eq:expt}) has exactly the form of a classical expectation value.  So why not treat $F_\psi(p,x,t)$ as a probability distribution?  

Unfortunately it is not difficult  to find situations in which $F_\psi(p,x,t)$ becomes negative so it cannot be a true probability.  Nevertheless it can be used to calculate correct expectation values, hence the term ``quasi-probability function".  However the appearance of negative probabilities  is not a satisfactory feature in a theory aimed at generalising ordinary statistical methods to cover quantum phenomena.  Of course Moyal was well aware of this difficulty.  He was also aware that Dirac \cite{pd42} had earlier discussed the appearance of negative probabilities in the Klein-Gordon equation, arguing that negative probabilities should not simply be ``considered as nonsense".  ``They are well defined concepts mathematically, like a negative sum of money", an idea that was developed further in a paper by Bartlett \cite{msb45}.

The question of negative probabilities was later taken up again by Sudarshan \cite{ecgs} and more recently by Feynman \cite{rf} who also argued that negative values were quite acceptable provided the negative values do not show up in any observable situation.  The method can give a convenient way of calculating expectation values which may prove more difficult to evaluate using standard techniques \cite{pcfz83}.  

In another approach Bohm and Hiley \cite{boh81} have shown how these negative probabilities can be circumvented if desired, but this leads to a different theory.  But notice the debate about negative probabilities only arises if we attempt to interpret $F_\psi(p,x,t)$ as a probability density in equation (\ref{eq:expt}).  Before dismissing the approach out of hand, let us  look at the meaning of this term from two different points of view.

\subsection{Expectation values}

We have seen how the expectation value of an operator can be written in the form (\ref{eq:expt}) whereas the usual form of the expectation value is written as
\begin{eqnarray*}
\langle \hat A(t)\rangle &=&\int\langle \psi(t)|x'\rangle\langle x'|\hat A|x''\rangle\langle x''|\psi(t)\rangle dx'dx''\\
&=&\int A(x',x'',t)\rho_\psi(x',x'',t)dx'dx''
\end{eqnarray*}
Let us now change to the coordinates 
\begin{eqnarray*}
x'=x-y/2;\quad\quad x''=x+y/2
\end{eqnarray*}
so that
\begin{eqnarray*}
\langle \hat A(t)\rangle=\int \langle x-y/2|\hat A|x+y/2\rangle\rho_\psi(x-y/2,x+y/2,t)dxdy		\label{eq:xy}
\end{eqnarray*}
Now define $F_\psi(p,x,t)$ by 
\begin{eqnarray*}
\rho_\psi(x-y/2,x+y/2,t)=\int F_\psi(p,x,t)e^{-iyp}dp.
\end{eqnarray*}
Substituting into equation (\ref{eq:xy}) and rearranging, we find
\begin{eqnarray*}
\langle \hat A(t)\rangle=\int\left[\int\langle x-y/2|\hat A(t)|x+y/2\rangle e^{-iyp}dy\right]F_\psi(p,x,t)dpdx.
\end{eqnarray*}
Defining the square bracket to be $a(p,x,t)$, we find
\begin{eqnarray*}
\langle \hat A\rangle=\int a(p,x,t)F_\psi(p,x,t)dpdx.
\end{eqnarray*}
which is identical to equation (\ref{eq:expt}).  This result shows quite clearly that all we have done is to change  coordinates and rearranged the standard expression for the expectation value of an operator $\hat A$.  As yet we still have not identified the exact meaning of the variables $(p,x)$.

\subsection{The Wigner Distribution and the Density Matrix.}

So far we have started from a definition of the characteristic function $M_\psi(\tau, \theta,t)$ and introduced a pair of unidentified variables $(p, x)$ through a Fourier transformation (\ref{eq:4}).  It is often assumed that these variables refer to the position and momentum of a particle but is this actually the case?
	
  Following on from the paper of Takabayasi \cite{tak}, Bohm and Hiley \cite{boh81} showed that one can construct a symplectic phase space from a pair of points in a configuration space.  To do this we start with a density matrix written in the form $\rho(x,x^{\prime})$, thus regarding it as a two point function in {\em configuration space}.  By writing this density matrix in the form
\begin{eqnarray}
\rho_\psi(x,x^{\prime},t)=\psi(x,t)\psi^{*}(x^{\prime},t)=\frac{1}{2\pi}\int\int\phi(p,t)e^{ixp}\phi^{*}(p^{\prime},t)e^{-ix^{\prime}p^{\prime}}dpdp^{\prime}\nonumber
\end{eqnarray}
and introducing new co-ordinates
\begin{eqnarray}
X=(x+x^{\prime})/2;  \hspace{0.3cm}  \tau=x-x^{\prime};  \hspace{0.3cm}
\mbox{and}  \hspace{0.3cm}   P=(p+p^{\prime})/2;  \hspace{0.3cm}  \theta=p-p^{\prime}  		\label{eq:8}
\end{eqnarray}										
the density matrix can be transformed into
\begin{eqnarray}
\rho_\psi(X,\tau,t)=\frac{1}{2\pi}\int\int\phi(P-\theta/2,t)e^{i\theta X}\phi^*(P+\theta/2,t)e^{i\tau P}dPd\theta.  \nonumber
\end{eqnarray}
We can now write this equation in the compact form
\begin{eqnarray}
\rho_\psi(X,\tau, t)=\int F_\phi(X,P,t)e^{i\tau P}dP	\label{eq:9}
\end{eqnarray}										
where
\begin{eqnarray}
F_\phi(P,X,t)=\frac{1}{2\pi}\int\phi^{*}(P+\theta/2,t)e^{iX\theta}\phi(P-\theta/2,t)d\theta.
\nonumber
\end{eqnarray}
Or taking the inverse Fourier transform we have
\begin{eqnarray}
F_\psi(P,X,t)=\frac{1}{2\pi}\int\psi^{*}(X-\tau/2,t)e^{-iP\tau}\psi(X+\tau/2,t)d\tau
\nonumber
\end{eqnarray}
which, of course, is just the Wigner distribution (\ref{eq:4}). 
 
Thus we see that the coordinates used in the Moyal distribution, $F_\psi(P,X,t)$, are {\em not} the position and momentum of a single particle, rather $X$ and $P$ are the mean position and momentum of what could be taken to be a cell in phase space.  But we are using the density matrix to describe a single particle, not an ensemble of particles.   Thus our single particle is not represented by a point but as an extended region or `blob' in phase space.  In some earlier papers  \cite{hil01} \cite{hil03} I tried to develop a quantum dynamics based on an evolution of a primitive cellular structure in phase space with limited success. 

To make more progress with these ideas it is necessary first to realise that the space we are talking about has a symplectic structure.  To see this, let us return to equation (\ref{eq:hesgen}) which will be recognised as a generator of the Heisenberg group.  These generators satisfy
\begin{eqnarray*}
\widehat S(\alpha, \beta)\widehat S(\alpha', \beta')=e^{i\sigma}\widehat S(\alpha', \beta')\widehat S(\alpha, \beta)
\end{eqnarray*}
where $\sigma = (\alpha'\beta-\alpha\beta')$, an invariant antisymmetric bilinear form which implies that the $(\alpha, \beta)$ space is a symplectic space.  This, in turn, implies that the $(X,P)$ space is also a symplectic space which is necessary if we are to call the $(X,P)$ space a `phase space'.

  This now opens up a more fruitful approach that has been proposed by de Gosson \cite{mdg12}.  It means that we are building quantum mechanics on a symplectic space, a space upon which classical mechanics is also built.  Thus we see that classical mechanics and quantum mechanics are different structures built on the same underlying space, which means that this approach offers the possibility of giving a clearer understanding of the relation between the two.  Symplectic spaces have topological invariants called symplectic capacities, which have shown to be intimately related to the uncertainty principle \cite{gos2}. In this respect it seems as if classical mechanics already contains a shadow of the uncertainty principle.

\subsection{The Meaning of the Weyl Smbol $a(p,x,t)$}

Having clarified the meaning of $F_\psi(p,x,t)$ \footnote{We will follow convention and revert to lower case $X$ and $P$.} let us now examine the 
 precise physical meaning of the Weyl symbol $a(p,x,t)$ introduced in equation (\ref{eq:expt}).  

  Let us start with the usual definition of the mean value of the operator $\hat A$,
  \begin{eqnarray*}
\langle \hat A\rangle=\langle \psi(t)|\hat A|\psi(t)\rangle=\int\langle\psi(t)|x'\rangle\langle x'|\hat A|x''\rangle\langle x''|\psi(t)\rangle dx'dx''\\=\int\langle x'|\hat A|x''\rangle \rho(x',x'',t)dx'dx''\hspace{0.65cm}
\end{eqnarray*}
Let us now change coordinates using
$x'=x-\tau/2$ and $ x''=x+\tau/2,$
then
\begin{eqnarray*}
\langle \hat A\rangle=\int\langle x-\tau/2|\hat A|x+\tau/2\rangle \rho(x-\tau/2,x+\tau/2,t)dxd\tau
\end{eqnarray*}
Now write $\rho(x-\tau/2,x+\tau/2,t)=\int F_\psi(p,x,t)e^{ip\tau}dp$, and we find
\begin{eqnarray*}
\langle \hat A\rangle=\int\langle x-\tau/2|\hat A|x+\tau/2\rangle  e^{-ip\tau}d\tau\;[ F_\psi(p,x,t)dpdx],
\end{eqnarray*}
which becomes equation (\ref{eq:expt}) if we identify
\begin{eqnarray*}
a(p,x,t)=\int\langle x-\tau/2|\hat A(t)|x+\tau/2\rangle  e^{-ip\tau}d\tau.
\end{eqnarray*}

Thus we see that $a(p,x,t)$ is a transition probability amplitude integrated over the `blob' at the mean position $x$ when the blob is moving with mean momentum $p$.  The Weyl symbol $a(p,x,t)$ is sometimes called ``classical observable associated with the observable $\hat A$", but I find that misleading since there  is very little that is classical about $a(p,x,t)$.  It is  a transition probability amplitude.  In section (\ref{sec:MA}) we will show that these symbols combine under a non-commutative  multiplication rule, $a(p,x,t)\star b(p,x,t)$, emphasising again that they are definitly not classical functions.

Let us conclude this section by pointing out that $F_\psi(p,x,t)$ is the Weyl symbol of the density operator.  To show this let us write equation (\ref{eq:4}) in a slightly different form
\begin{eqnarray*}
F_\psi(p,x,t)=\frac{1}{2\pi}\int e^{-ip\tau}\langle x-\tau/2,t|\;(|\psi\rangle\langle \psi|)\;|x+\tau/2,t\rangle d\tau.
\end{eqnarray*}
Notice $(|\psi\rangle\langle\psi|)$ is just the density operator, $\hat \rho$, for a pure state, so that we can write
\begin{eqnarray}
\rho_\psi(p,x,t)=\frac{1}{2\pi}\int e^{-ip\tau}\langle x-\tau/2,t|\hat \rho|x+\tau/2,t\rangle d\tau.
\label{eq:5}			
\end{eqnarray}
Thus clearly demonstrating that 
 the probability distribution $F_\psi(p,x,t)[:=\rho_\psi (p, x,t)]$ is simply the Weyl symbol for the density matrix for a pure state in the $(p,x,t)$ representation.  Note there is no reason why a density matrix should stay positive.  The positivity condition is only desirable if $F_\psi(p,x,t)$ is to be regarded as a probability density.

\subsection{Moyal's contribution}

Since the mathematical structure originally introduced by von Neumann is identical to the one later used by Moyal, I would like to clarify the significance of Moyal's contribution. 

von Neumann left his results expressed in terms of parameters, $\alpha$, and $\beta$, whose physical meanings were unspecified and so the physical significance of the approach was unclear.   What Moyal noticed was that a simple Fourirer transformation will put the von Neumann expression $\langle\psi|\widehat S(\alpha, \beta)|\psi\rangle$  into a form that looked like a classical probability density in an $(p,x,t)$ phase space so that equation (\ref{eq:expt})  became a straight forward statistical average. 
This immediately suggest that $\widehat S(\alpha, \beta)$ could be regarded as a {\em characteristic operator}, a {\em quantum generalisation} of the characteristic function which plays a central role in the general theory of statistics.  Since equation (\ref{eq:expt}) looks like a classical average perhaps quantum mechanics is basically a statistical theory over a phase space, even though this seems to be called into question by the uncertainty principle\footnote{If $x_1$ and $x_2$ are conjugate points, then one can show $[\hat X,\hat P]=0$.}.

There are three things to notice about this generalisation
\begin{itemize}
\item The density matrix we start with is a {\em two point function} in configuration space.
\item  These two points are transformed into a {\em one point function} in the higher dimensional phase space, that is, a space where each point is now given the coordinate $(p,x)$.
\item  The density matrix is applied to the description of a single particle, rather than an ensemble of particles.
\end{itemize}

What does this all mean for the behaviour of a single particle?  It suggests that in the quantum domain we can no longer describe the particle by a local entity, such as a precisely localised `small rock' with sharp position $x$ and a sharp momentum $p$.  Its description requires a region of space in order to describe its behaviour.  That is,  it is behaving like an extended `blob' \cite{mdg12} specified by a mean position and a mean momentum. To get a deeper understanding of what all this means, we must explore further the mathematical structure of the Moyal algebra.

\subsection{The Moyal Algebra}\label{sec:MA}

Let us begin by exploring the formal mathematical structure of the Moyal algebra a little further.  What we  will show is that the phase space is a {\em non-commutative phase space}. This has profound implications for the underlying symplectic geometry  (See de Gosson \cite{gos2}, \cite{gos1}). 

von Neumann showed that to completely reproduce the results of quantum mechanics, it is necessary to introduce a non-commutative star product, which in the $(p,x)$ representation, is defined by
 \begin{eqnarray}
 a(p,x)\star b(p,x)=\int\int e^{2i(p\eta-x\xi)}a(p-\xi,x-\eta)\bar b(\xi,\eta)d\xi d\eta,			\label{eq:starvn}		
 \end{eqnarray}
 where
 \begin{eqnarray*}
 b(p,x)=\int\int e^{2i(p\eta-x\xi)}\bar b(\xi,\eta)d\xi d\eta.
 \end{eqnarray*}
 Although the product was first introduced by von Neumann, it is now called the Moyal star product, since it was Moyal that showed it could be written as a formal series
\begin{eqnarray}
a(p,x)\star b(p,x)= a(p,x) \exp\left[\frac{i\hbar}{2}\left(\overleftarrow{\frac{\partial}{\partial x}}\overrightarrow{\frac{\partial}{\partial p}}-\overrightarrow{\frac{\partial}{\partial x}}\overleftarrow{\frac{\partial}{\partial p}}\right)\right]b(p,x).  \label{eq:star}		
\end{eqnarray}
This form is extremely useful for cases where the $a(p,x)$ and $b(p,x)$ are finite polynomials.  For example it is trivial to show that
\begin{eqnarray*}
x\star p -p\star x=i\hbar.
\end{eqnarray*}
showing how the famous Heisenberg defining relation appears in this phase space.  As a passing remark, it should be noted that 
for non-polynomial functions, product rule (\ref{eq:starvn}) should be used to avoid convergence problems.

The star product enables us to define two types of bracket; the Moyal bracket defined by
\begin{eqnarray*}
\{a,b\}_{MB}=\frac{a\star b-b\star a}{i\hbar},
\end{eqnarray*}
and the Baker bracket \cite{bak}  (or Jordan product) defined by
\begin{eqnarray*}
\{a,b\}_{BB}=\frac{a\star b+b\star a}{2}.
\end{eqnarray*}
Clearly the Moyal bracket replaces the quantum operator commutation relations $[\hat A, \hat B]$.  

The nature of the $\star$-product means that the Moyal bracket will, in general, be a power series in $\hbar$.  If we  retain only the terms to $O(\hbar)$, we find 
\begin{eqnarray*}
\mbox{Moyal bracket}\rightarrow \mbox{Possion bracket}.
\end{eqnarray*}
 While in the case of the Baker bracket we find
\begin{eqnarray*}
\mbox{Baker bracket}\rightarrow \mbox{simple commutative product}.
\end{eqnarray*}
Thus we find that the Moyal approach contains classical physics as a limiting case.  There is no need to look for a correspondence between  commutator brackets and Poisson brackets, a process which fails as is demonstrated by the Groenwald-van Hove theorem \cite{vgss} in their well known `no-go' theorem.

Furthermore there is no need to introduce the notion of decoherence as a fundamental process to obtain the classical limit.  This does not mean that decoherence has no role in quantum physics.  It plays a vital role in real experiments where noise and other thermal processes enter to destroy quantum interference, but destroying the interference does not return us to the classical formalism involving Poisson brackets.

\section{Moyal Dynamics}

\subsection{Moyal's Original Equation}

In order to discuss the dynamical time evolution of the quasi-probability density $F_\psi(p,x,t)$, Moyal starts from the Heisenberg equation of motion for $M_\psi(\tau, \theta,t)$ which he writes as
\begin{eqnarray*}
i\frac{\partial  M_\psi(\tau,\theta,t)}{\partial t}=\int \psi^*(x,t)[\widehat H, \widehat M]\psi(x,t)dx,
\end{eqnarray*}
where $\widehat H$ is the Hamiltonian of the system. After some working, the details of which can be found in \cite{moy}, we find
\begin{eqnarray}
\frac{\partial M_\psi(\tau,\theta,t)}{\partial t}=i\int[H(p+\theta/2,x-\tau/2)-H(p-\theta/2,x+\tau/2)]\nonumber\\\times F_\psi(p,x.t)e^{i(\tau p+\theta x)}dpdx.\hspace{2cm}		\label{eq:MHE}		
\end{eqnarray}
By using
\begin{eqnarray*}
M_\psi(\tau,\theta,t)=F_\psi(\eta, \xi,t)e^{i(\tau\eta+\theta\xi)}d\eta d\xi,
\end{eqnarray*}
equation (\ref{eq:MHE}) can be put in the form
\begin{eqnarray*}
\frac{\partial F_\psi(p,x,t)}{\partial t}+2F_\psi(p,x,t)\sin\frac{\hbar}{2}[\overleftarrow\partial_x\overrightarrow \partial_p-\overrightarrow\partial_x\overleftarrow \partial_p]H(p,x)=0.
\end{eqnarray*}
This equation can finally be written as
\begin{eqnarray}
\frac{\partial F_\psi(p,x,t)}{\partial t}+\{F_\psi(p,x,t),H(p,x)\}_{MB}=0.
\label{eq:liou}
\end{eqnarray}										
Thus the Moyal bracket plays a key role  in the time evolution of the system. 

As we have already remarked, this equation becomes the  classical Liouville equation if the Moyal bracket is expanded as a series in powers of $\hbar$, retaining only those terms of order $\hbar^2$.  Thus the mathematical structure of classical mechanics emerges as an approximation when the higher order terms in $\hbar$ become negligible.  However this is only a part of the story and we need to consider the role played by the often neglected Baker bracket in the dynamics.

\subsection{The Key Role played by the Wigner Density Function}

From equation (\ref{eq:4}) we see that in the Moyal approach, it is the Wigner density function, not the wave function, that plays a key role.  Recall this function is a transition probability, and not a state function, a point  recognised by Groenewold \cite{hg83} who called it an `ensemble operator', using the word `ensemble'  even when the formalism is applied to a single quantum system.  

Why use  a statistical concept to describe a single system?  Groenewold was quite clear. There is a certain ambiguity in the information and control of a single quantum system that seems to necessitate a statistical description in quantum mechanics. This is a feature that has long been recognised and attempts to remove this ambiguity has lead to the quest for `hidden variables'.   Groenewold left open the question as to whether we can ultimately find a less ambiguous description.  

As no less ambiguous description has, in fact, been found to date, my own conclusion is that, in quantum processes, we have reached the limits of the paradigm that allows us to make a sharp, unambiguous  separation between subject and object.  In this I agree with Bohr \cite{nb61}.  This claim may come as a surprise to many since I have been deeply involved in the Bohm approach.  It is true that this approach originally started out as an attempt to make such an unambiguous description but the appearance of an additional term, the quantum energy potential in the conservation of energy equation (see equation \ref{eq:energycons}), blurs this distinction as this term contains irreducible information about relation between subject and object, as was discussed in detail by Bohm and Hiley \cite{dbbh93}.

 My own position and indeed that of David Bohm, from the sixties when we first became colleagues, was to realise the inadequacy of a descriptive form that depends on objects-in-interaction.  Already in his book ``Causality and Chance in in Modern Physics" \cite{db57}, Bohm argued that we had reached the limits of the mechanical paradigm and something more radical was needed. Indeed  Bohm himself had already started to explore a process oriented description in the late fifties, publishing a key paper on the subject in the early sixties  \cite{db65}.  I have also explained my own reasons for exploring such an approach elsewhere so I will not repeat them here \cite{bhdist}. 
 
 Although we have motivated the Moyal approach in this paper by using the concept of a wave function, one can start by taking the Wigner-Moyal density function (\ref{eq:4}) as basic.  But as we have already pointed out, this is a transition probability amplitude, the description is well suited to a process approach, although we will not develop this line of reasoning here.  
 
 Again as we have shown earlier, this density function is a specific representation of the standard density matrix, which is a generalisation of the wave function, introduced to describe mixed states.  Thus for example the more general mixed state is described by $\rho(x,t)=\sum c_n \psi_n(x,t)$ with $\sum|c_n|^2=1$.  Notice, however, we have again made use of the notion of a wave function.  But the statistical distribution function $F_\psi(p,x,t)$ is potentially more general since we can start with the general defining equation (\ref{eq:expt}) written in the form
 \begin{eqnarray}
 \langle \hat A(t)\rangle = \int\int a(p,x,t) F(p,x,t)dpdx,  \label{eq:expt1} 
 \end{eqnarray}
  If we want to adopt this position, then we must show exactly how the wave function emerges from this more general starting point. Fortunately Baker \cite{bak} has already done this in a paper whose importance has been almost forgotten.  He shows that if we use the relation 
  \begin{eqnarray} 
  \{F(p,x,t),F(p,x,t)\}_{BB}=F(p,x,t),	\label{eq:idempot}	
  \end{eqnarray}
   then we can always write
   \begin{eqnarray*}
   \int e^{ip\tau}F(p,x,t)dp=g^*(x-\tau/2,t)g(x+\tau/2,t),
   \end{eqnarray*}
allowing us to identify the function  $g(y,t)$ with what we traditionally call the `wave function'.  In other words when $F(p,x,t)$ is an idempotent satisfying (\ref{eq:idempot}), we can associate a  wave function with the system.  This function then enables us to use the Schr\"{o}dinger algorithm to calculate the statistical properties of the system in the way Bohr suggested.  The analogy with the density matrix is then quite clear since it is well known that if the density matrix is idempotent, it describes a pure state.  However we want to stress that in the Moyal approach, the wave function is a {\em derived} notion, a point that has been stressed by de Gosson \cite{mdg06}.

What is important from the point of view we are developing here is the idempotent nature of the density function.  As we have already indicated above we are replacing objects-in-interaction with the notion of a structure process introduced by Bohm \cite{db65}.  What, then, is an `object' in a process philosophy?  Since all is process, a system that persists like a `particle' must be something that keeps transforming into itself. The flame of a candel is something that continues to exist, yet the gasses giving rise to the flame are being replaced continuously provides a nice metaphore.  

In a detailed analysis of  the notion of `existence', Eddington \cite{ae} repudiates any metaphysical concept of existence like a `particle', and replaces it by a structural concept of existence which is mathematically defined by an idempotent\footnote{ The eigenvalues of an idempotent operator are 1 or 0, to exist or not to exist.}.  We are also adopting this idea and regarding objects as idempotents in the process, hence we have a straight forward way to account for the instability of quantum particles in general, as well as pair creation and annihilation.  Notice also a quantum object cannot be isolated as can a classical object.  They are a feature of the overall structure process.  Without the process, they do not exist.  Alter the overall process and the properties of the individual object change.  This then goes a long way to account for Bohr's insistence on a key notion of quantum phenomena, namely, the non-separability of observed system from the means of observation.

\subsection{The Complete Dynamics}

Let us return to the time evolution of the density function, $F_\psi(p,x,t)$, but now assuming our system to be in a pure state being specified by an idempotent density function.  We have seen how  the commutator in the Heisenberg equation of motion is replaced by the Moyal bracket, but this does not completely determine the the time evolution of the system as has indeed already been pointed out by Carruthers and Zachariasen \cite{pcfz83}. We must find the role played by the Baker bracket in this evolution.

Our algebra is non-commutative, so that we have to distinguish between left and right multiplication, between $H(p,x)\star F_\psi(p,x,t)$ and $F_\psi(p,x,t)\star H(p,x)$, where $H(p,x)$ is the Hamiltonian.  This means that we have two equations for the time development,
\begin{eqnarray}
H(x,p)\star F_\psi(x,p,t)=i(2\pi)^{-1}\int e^{-i\tau p}\psi^*(x-\tau/2)\overrightarrow\partial_t\psi(x+\tau/2)d\tau  \label{eq:LM}	
\end{eqnarray}
and
\begin{eqnarray}
F_\psi(x,p,t)\star H(x,p)=i(2\pi)^{-1}\int e^{-i\tau p}\psi^*(x-\tau/2)\overleftarrow\partial_t\psi(x+\tau/2)d\tau.  \label{eq:RM}	
\end{eqnarray}
Now if we subtract these two equations, we immediately obtain equation (\ref{eq:liou}).  However if we add the two equations we obtain
\begin{eqnarray}
\{H,F\}_{BB}=i(2\pi)^{-1}\int e^{-i\tau p}\left[\psi^*(x-\tau/2)\overleftrightarrow\partial_t\psi(x+\tau/2)\right]d\tau. 	\label{eq:bb}
\end{eqnarray}  
We have introduced a condensed notation which is well known in quantum field theory\footnote{The distinction between left and right multiplication is necessary even in conventional quantum field theory when one deals with the Pauli and Dirac particles.  The double arrow symbol (\ref{eq:2way}) is used for the energy term, for example, in the Lagrangian for the Dirac field \cite{pr81}.  It is therefore not surprising that equation (\ref{eq:bb}) involves energy.}. viz,
\begin{eqnarray}
\psi^*\overleftrightarrow \partial_t \psi=\psi^*(\partial_t\psi)-(\partial_t\psi^*)\psi.		\label{eq:2way}	
\end{eqnarray}
We can quickly get an idea as to what the RHS means if we choose an energy eigenstate, $\psi(x, \tau,t)=\phi(x, \tau)e^{iEt}$.  We  find
\begin{eqnarray*}
i(2\pi)^{-1}\int e^{-i\tau p}\left[\psi^*(x-\tau/2)\overleftrightarrow\partial_t\psi(x+\tau/2)\right]d\tau =-2EF_\psi(p,x,t).
\end{eqnarray*}
So clearly the Baker bracket has something to do with the energy of the system.  Therefore let us condense the notation by writing
\begin{eqnarray*}
i(2\pi)^{-1}\int e^{-i\tau p}\left[\psi^*(x-\tau/2)\overleftrightarrow\partial_t\psi(x+\tau/2)\right]d\tau=-{\cal E}(p,x,t),
\end{eqnarray*}
so that we can write equation (\ref{eq:bb}) in the form
\begin{eqnarray*}
{\cal E}(p,x,t)+ \{H,F\}=0
\end{eqnarray*}

Notice also that if we use a cross-Wigner function which employs two different energy eigenfunctions, then the Baker bracket measures the mean energy of the two eigenstates.  Such an equation was first introduced by Dahl \cite{jpd83}  to supplement the Liouville equation in order to obtain a complete specification of the energy eigenstates  of molecules \cite{pcfz83}.

The full significance of equation (\ref{eq:bb}) is still not obvious from the form of the the RHS of equation (\ref{eq:bb}) but a further insight can be found by going to the limit $O(\hbar^2)$.  After some work, including writing $\psi(x,t)=R(x,t)\exp[iS(x,t)]$, we find
\begin{eqnarray*}
\{H,F\}_{BB}=-2(\partial_tS) F+O(\hbar^2)
\end{eqnarray*}
From the definition of the Baker bracket we find that in this limit equation (\ref{eq:bb}) becomes
\begin{eqnarray*}
\frac{\partial S}{\partial t}+H=0
\end{eqnarray*}
This will immediately be recognised as the classical Hamilton-Jacobi equation. Thus equation (\ref{eq:bb}) is the quantum generalisation of the classical Hamilton-Jacobi equation.  In the next section we will relate this to the quantum Hamilton-Jacobi equation introduced by Bohm and Hiley \cite{dbbh93}.

\subsection{Summary of time development equations}

Since we are dealing with a non-commutative structure, we have to distinguish between left multiplication in equation (\ref{eq:LM}) and right multiplication in equation (\ref{eq:RM}).  These equations give rise to {\em two} time development equations, the first being equation (\ref{eq:liou}) was written in the form
\begin{eqnarray}
\frac{\partial F_\psi(p,x,t)}{\partial t}+\{F_\psi(p,x,t),H(p,x)\}_{MB}=0	\label{eq:LE}		
\end{eqnarray}
The second equation is 
\begin{eqnarray}
{\cal E}(p,x,t)+\{H,F\}_{BB}=0 	\label{eq:MQHJ}	
\end{eqnarray}

Let us emphasise again that the Moyal bracket is equivalent to the commutator
\begin{eqnarray*}
\{H,F\}_{MB}\quad\leftrightarrow\quad [\hat H,\hat\rho]_-
\end{eqnarray*}
while the Baker bracket is the Moyal equivalent of the anti-commutator 
\begin{eqnarray*}
\{H,F\}_{BB}\quad\leftrightarrow\quad [\hat H, \hat \rho]_+
\end{eqnarray*}

The operator equation corresponding to equation (\ref{eq:LE}) is the quantum Liouville equation
\begin{eqnarray*}
i \frac{\partial \hat\rho_\psi(x,t)}{\partial t}+[\hat H(x), \hat\rho_\psi(x,t)]_-=0
\end{eqnarray*}
while equation (\ref{eq:MQHJ}) has the operator equivalent is
\begin{eqnarray}
2\partial_tS\hat\rho+[\hat \rho,\hat H]_+=0.		\label{eq:bb0}	
\end{eqnarray}
These operator equations are obtained by taking conditional expectation values, i.e. integrating over $p$.
For an alternative derivation of these operator equations see Brown and Hiley \cite{bro}.

Equation (\ref{eq:bb0}) is very significant for those with interest in the Bohm interpretation.  Let us choose the Hamiltonian to be $\hat H = \hat p^2/2m +V$, then equation (\ref{eq:bb0}) becomes
\begin{eqnarray}
\partial_t S(x,t)+\frac{1}{2m}[\partial_x S(x,t)]^2+Q(x,t)+V(x,t)=0	\label{eq:energycons}	
\end{eqnarray}
 where $Q(x,t)= \nabla^2R(x,t)/2mR(x,t)$, the quantum potential energy.
Equation (\ref{eq:energycons}) is just the quantum Hamilton-Jacobi equation, the real part of the Schr\"{o}dinger equation used in the Bohm approach.  A different way of showing that the equations used in the Bohm model can be derived from the Moyal algebra will be found in Hiley \cite{bh11}. But this is not simply another way of deriving the Bohm results.  It is vital for generalising the Bohm model to include relativistic spin as shown by Hiley and Callaghan \cite{bhbc11}.

\section{Conclusions.}

We have shown here that the algebraic structure of the Moyal approach \cite{moy} is isomorphic to the algebraic approach introduced in a much earlier paper by von Neumann \cite{vn32}.   Thus Moyal uses an algebra without modification that is at the very heart of the quantum formalism. However von Neumann offers no interpretation of the formalism at all, whereas Moyal provides a specific physical interpretation in terms of a statistical approach based on probabilities in a generalised phase space. 
Unfortunately the interpretation leads to a problem, namely, that  probabilities can take on negative values and this has left the general impression that the Moyal algebra is inadequate in some way, in spite of discussions by Dirac \cite{pd42}, amongst others \cite{msb45}, \cite{rf}, that accommodate negative probabilities.  It should be noted that the Moyal-von Neumann algebra is sound quantum mechanics and it is only the interpretation that is blighted.

We have also confirmed Baker's \cite{bak} original conclusion  that the Wigner distribution is simply  the density matrix expressed in a special representation involving the mean position of a pair of points in configuration space.  The Wigner-Moyal transformation then enables us to construct a six-dimensional non-commutative phase space for a single particle \cite{boh81}.  In this space we see that the density matrix (Wigner function) describes, not a point particle, but an extended structure, the coordinates $(p,x)$ describe the mean momentum and mean position of the structure.  de Gosson has called this  structure, the quantum blob \cite{gos1}.

The `blob' structure was hinted at by Baker \cite{bak}, who argued that the Wigner-Moyal formalism introduces a kind of ``smeared-out'' projection operator for a region in phase space.  However recently de Gosson \cite{gos1} \cite{gos2} has taken this further by pointing out that this structure can be related to  a rich topological structure that underlies symplectic spaces as demonstrated by Gromov's \cite{gro} ``no squeezing theorem''.  This theorem  shows that there are areas of phase space involving pairs of conjugate co-ordinates that cannot be reduced in size even under a classical symplectomorphism. This is a kind of classical harbinger of the quantum uncertainty principle.  What quantum mechanics does is to introduce a minimum value, $\hbar$, for this area.  More generally  these `quantum blobs' can be discussed formally in terms of the notion of a symplectic capacity.  Further work along these lines has been reported by de Gosson \cite{gos3}.

As a final remark I would like to connect the ideas outlined in this paper to the more radical ideas I have been pursuing elsewhere \cite{hil03},\cite{boh06}.
Strictly speaking the Wigner function is a transition probability function which suggests the approach be based on the fundamental notion of process.  Without a basic process philosophy, it would be difficult to provide a physical motivation for giving primary relevance to the non-commutative structure.  It should be noted that within this structure, not only do we have quantum mechanics, but we also have classical mechanics arising naturally in the limit of order $\hbar$.  Thus this algebraic discription provides a natural approach to both quantum and classical physics so that there is no need to call on the notion of  `decoherence' to explain the emergence of the classical world.

\section{Acknowledgments}
I would like to thank Fabio Frescura for many fruitful discussions and for drawing my attention to the von Neumann paper in the first place.  I would also like to thank Maurice de Gosson for his help in my struggle to understand the deep topological structures that lie in symplectic geometry.  I would also like to thank the members of the TPRU for their many helpful comments as this work unfolded.


\begin{thebibliography}{99}

\bibitem{bak}   Baker, G. A., Jr., Formulation of Quantum Mechanics Based on the Quasi-Probability Distribution Induced on Phase Space, {\em Phys. Rev.}, {\bf 109}, (1958), 2198-2206.

\bibitem{msb45} Bartlett, M. S., Negative Probabilities, {\em Proc. Cambridge Phil. Soc.}  {\bf 41}, (1945), 71-73.

\bibitem{db57}  Bohm,  D.,  {\em Causality and Chance in Modern Physics}, Routledge \& Kegan Paul, London, (1957).

\bibitem{db65} Bohm, D.,  Space, Time, and the Quantum Theory Understood in Terms of Discrete Structural Process, {\em Proc. Int. Conf. on Elementary Particles}, Kyoto, pp. 252-287, (1965)

\bibitem{boh81}   Bohm, D. and Hiley, B. J., On a Quantum Algebraic Approach to a Generalised Phase Space,  {\em Found. Phys}., {\bf 11}, (1981), 179-203. 

\bibitem{boh83} Bohm, D. and Hiley, B. J.,    Relativistic Phase Space Arising out of the Dirac Algebra,  in {\em Old and New Questions in Physics and Theoretical Biology}, Ed., A. van der Merwe,  pp. 67-76,  Plenum, 1983.

\bibitem{dbbh93} Bohm, D. and Hiley, B. J., {\em The Undivided Universe: An Ontological Interpretation of Quantum Mechanics}, Routledge, London, 1993.

\bibitem{boh06}  Bohm, D.. Davies, P. G., and Hiley, B. J.,  Algebraic Quantum Mechanics and Pre-geometry,  in AIP Conference Proceedings, {\bf 810}, {\em Quantum Theory: Reconsideration of Foundations--3, V\"{a}xj\"{o}, Sweden, 2005},  ed. Adenier, G.,  Khrennikov, A.,  Nieuwenhuizen, Theo., pp. 314-324 AIP, New York, 2006.

\bibitem{nb61} Bohr, N.,   {\em Atomic Physics and Human Knowledge}, Science Editions, New York, 1961.

\bibitem{bro}  Brown, M. R. and Hiley,  B. J.,  Schr\"{o}dinger revisited: an algebraic approach,  {\em quant-ph/0005026}.

\bibitem{pcfz83} Carruthers, P. and Zachariasen, F., Quantum collision theory with phase-space distributions, {\em Rev. Mod. Phys.}, {\bf 55}, (1983), 245-85.

\bibitem{ac90}  Crumeyrolle A.,   {\em Orthogonal and Symplectic Clifford Algebras:  Spinor Structures}, Kluwer, Dordrecht, 1990.

\bibitem{dem} Demaret, J., Heller, M. and Lambert,  D., Local and Global Properties of the world,   {\em Foundations of Science}, {\bf 2}, (1997),137-176.

\bibitem{jpd83} Dahl, J. P., Dynamical Equations for the Wigner Functions, in {\em Energy Storage \& Redistribution in Molecules}, Ed., Hinze, J. pp. 557-71, Plenum, New York, 1983.

\bibitem{pamd}  Dirac, P. A. M.,  A new notation for quantum mechanics, {\em Mathematical Proceedings of the Cambridge Philosophical Society}, {\bf 35}, (1939),  416-418.   doi: 10.1017/S0305004100021162

\bibitem{pd42} Dirac, P. A. M., The Physical Interpretation of Quantum Mechanics, {\em Proc. Roy. Soc.,} {\bf 180}, (1942), 1-40.

\bibitem{ae}  Eddington,  A. S.,    {\em The Philosophy of Physical Science}, p.162, Ann Arbor Paperback, University of Michigan Press, Michigan, 1958.

\bibitem{rf} Feynman, R. P., Negative Probability, in {\em Quantum Implications: Essays in Honour of David Bohm},  ed. by Hiley, B. J. and D.Peat, pp. 235-54,  Routledge \& Kegan Paul, 1987.

\bibitem{fre80} Frescura,  F. A. M. and Hiley, B. J.,  The Implicate Order, Algebras, and the Spinor,  {\em Found. Phys.}, {\bf 10}, (1980),  7-31.

\bibitem{fre84} Frescura,  F. A. M. and Hiley, B. J., Algebras, Quantum Theory and Pre-Space,  {\em Revista Brasilera de Fisica, Volume Especial,  Os 70 anos de Mario Schonberg},  (1984), 49-86.

\bibitem{gla}  Glauber, R. J., Quantum Optics and Electronics in {\em  Les Houches Lectures 1964}, ed De Witt et al, p. 139, Gordon and Breach, New York, 1965.

\bibitem{gos2}  de Gosson, M., Phase space quantization and the uncertainty principle. {\em Phys. Lett.}, {\bf A 317}, (2003), 365-369.

\bibitem{gos1} de Gosson, M.,Uncertainty Principle, Phase Space Ellipsoids and Weyl Calculus, in {\em Operator Theory: Advances and Applications}, {\bf 164}, 121-132, Birkh\"{a}user, Basel, Switzerland 2006.

\bibitem{gos3} de Gosson, M., Symplectic non-squeezing theorems, EBK quantization and quantum uncertainty, To be published.

\bibitem{mdg06} de Gosson, M., {\em Symplectic Geometry and Quantum Mechanics}, Birkh\"{a}user Verlag, Basel, 2006.

\bibitem{mdg12} de Gosson, M., Quantum Blobs, {\em  Found. Physics},  (2012),   1-18.  DOI 10.1007/s10701-012-9636-x

\bibitem{jgbjv95}  Gracia-Bond\'{i}a, J. M. and V\'{a}rilly, J. C.,  From geometric quantization to Moyal quantization, {\em J. Math. Phys.},  {\bf 36}, (1995), 2691-2701. 

\bibitem{hg83}  Groenewold, H. J., Pruned Quantum Theory, {\em Phys. Rep.} {\bf 98}, (1983), 343-365.

\bibitem{gro} Gromov, M., Pseudoholomorphic curves in symplectic manifolds, {\em Invent. Math.}, {\bf 82}, (1985), 307-47.

\bibitem{vgss}  Guillemin, V. and Sternberg, S.,  {\em Symplectic Techniques in Physics}, Cambridge University Press, Cambridge, 1984.

 \bibitem{hil01}  Hiley, B. J., Towards a Dynamics of Moments: The Role of Algebraic Deformation and Inequivalent Vacuum States, {\em Correlations}, ed. K. G. Bowden, Proc. ANPA {\bf23}, (2001), 104-134.
 
 \bibitem{hil01a}  Hiley, B. J., A Note on the Role of Idempotents in the Extended Heisenberg Algebra, {\em Proc.  Int, Meeting, ANPA}  {\bf 22}, 107-121, Cambridge, 2001.

 \bibitem{hil03}  Hiley, B. J., Phase Space Descriptions of Quantum Phenomena, Proc. Int. Conf. Quantum Theory, {\em Proc. Int. Conf. Quantum Theory: Reconsideration of Foundations},  {\bf 2}, 267-86, ed. Khrennikov, A., V\"{a}xj\"{o} University Press, V\"{a}xj\"{o}, Sweden, 2003.
 
 \bibitem{bhdist}  Hiley, B. J., Process, Distinction, Groupoids and Clifford Algebras: an Alternative View of the Quantum Formalism, in {\em New Structures for Physics}, ed Coecke, B., Lecture Notes in Physics, vol. 813, pp. 705-750, Springer, 2011.
 
 \bibitem{bhbc11} Hiley, B. J. and Callaghan, R. E.,  Clifford Algebras and the Dirac-Bohm Quantum Hamilton-Jacobi Equation. {\em Foundations of Physics}, {\bf 42}, (2012), 192-208.
 
 \bibitem{bj12}  Hiley, B. J., Weak Values:Approach through the Clifford and Moyal Algebras,  {\em J. Phys.: Conference Series,} {\bf 361}, (2012), 012014.   Doi: 10.1088/1742-6596/361/1/012014.  quant-ph/1111.6536.
 
 \bibitem{bh11} Hiley, B. J.,  On the Relationship between the Wigner-Moyal and Bohm Approaches to Quantum Mechanics: A step to a more General Theory?   {\em Found. Phys.} {\bf 40}, (2010), 356-367. DOI 10.1007/s10701-009-9320-y
 
 \bibitem{soc12}  Hiley, B. J., The Wigner-Moyal Approach to Relativistic Particle with Spin, in preparation, 2012. 

 \bibitem{mor} Moran, M and Manton, J. H.,  in {\em Computational Noncommutative Algebra and Applications}, ed.  Byrnes, J. S., pp. 339-62, NATO Science Series, Kluwer Academic, 2004.
 
 \bibitem{amoy} Moyal, A., {\em  Maverick Mathematician: The Life an Science of J. E. Moyal}, Australian National University E Press, Canberra, Australia, 2006. 
 
 \bibitem{moy} Moyal, J. E., Quantum Mechanics as a Statistical Theory, {\em Proc. Cam. Phil. Soc.}, {\bf 45}, (1949), 99-123.  
 
 \bibitem{von}  v. Neumann, J., Die Eindeutigkeit der Schr\"{o}dingerschen Operatoren, {\em Math. Ann.}, {\bf 104}, (1931), 570-87. 
 
 \bibitem{vn32} v. Neumann, J., {\em Mathematische Grundlagen der Quantenmechanik} Springer, Berlin, 1932.
 
 \bibitem{pr81} Ramond, P., {\em Field Theory: a Modern Primer}, Frontiers in Physics {\bf 51}, p. 48, Benjamin, Reading, Mass., 1981.
 
 \bibitem{ecgs}  Sudarshan, E. C. G., Structure of Dynamical Theories, in {\em Lectures in Theoretical Physics:  Brandeis Summer Institute,} {\bf 2},  (1961), 143-199.


\bibitem{tak}  Takabayasi, T., The formulation of Quantum Mechanics in terms of Ensemble in Phase Space, {\em Prog.  Theor. Phys.,} {\bf 11},  (1954), 341-74.

\bibitem{hw28}  Weyl, H.,  Quantenmechanik und Gruppentheorie, {\em Zeitschrift f\"{u}r Physik}, {\bf 46}, (1928), 1-46.

\bibitem{wig} Wigner, E.,  On the Quantum Correction for Thermodynamic Equilibrium, {\em Phys. Rev.}, {\bf 40}, (1932), 749-59.

\end{thebibliography}
\end{document}